\documentclass[12pt]{article}

\usepackage{caption}
\usepackage{epsfig}
\usepackage{color}
\usepackage{psfrag}
\usepackage{verbatim}
\usepackage{amsopn}
\usepackage{appendix}
\usepackage{dsfont,amsmath,amssymb,color,units,pstricks}
\usepackage{amsfonts}
\usepackage{graphicx}
\usepackage{natbib}
\usepackage{array}
\usepackage{lscape}
\usepackage{fullpage}
\usepackage{cancel}
\usepackage{url}
\usepackage{subcaption}
\usepackage{tikz}
\usepackage{ntheorem}
\usepackage{booktabs}
\usepackage{titlesec}

\allowdisplaybreaks

\newcommand{\ex}[1]{{\mathbb E}[#1]}

\newcommand{\std}[1]{{\mathrm{Std}}[#1]}

\newcommand{\be}{\begin{equation}}
\newcommand{\ee}{\end{equation}}
\newcommand{\bee}{\begin{equation*}}
\newcommand{\eee}{\end{equation*}}
\newcommand{\eq}[1]{Eq.~(\ref{#1})}

\theoremstyle{break}

\numberwithin{equation}{section}

\titleformat{\section} 
  {\normalfont\large\bfseries}
  {(\thesection) }
  {0.5em}
  {}
  []

\titlespacing*{\section}
  {0pt}
  {2.\baselineskip}
  {1\baselineskip}

\begin{document}

\unitlength = 1mm
\title{Diffusive in plain sight: An inconspicuous law of market impact}

\author{Julius F. Bonart}

\maketitle

\begin{abstract}
Decomposing market impact as the difference between realized and counterfactual returns, and requiring both to be diffusive, yields a structural identity that restricts admissible impact dynamics at the level of individual participants. This constraint implies the square-root law in the information-neutral regime and a crossover toward linear impact under strong informational coupling, consistent with empirical observations. It further implies that impact must adapt to each participant's order-flow statistics, with consequences for theories of optimal execution and no-arbitrage. In the information-neutral regime, cumulative impact is itself diffusive, providing a diagnostic that many propagator and latent-liquidity models fail to satisfy. Under this regime, market impact retains an undetermined all-pass degree of freedom that prevents its reduction to a pure surprise model. This residual freedom accommodates transient impact dynamics and can generate strictly positive impact costs.  
\end{abstract}


\section{Introduction}

A central empirical regularity of financial markets is that return autocorrelations are vanishingly small over time scales spanning more than five orders of magnitude, from sub-minute intervals to weeks~\citep{BouchaudBonartDonierGould2018}. 

This absence of predictability is generally viewed as structural: arbitrage eliminates persistent correlations. Price diffusivity is therefore adaptive and can coexist with strongly autocorrelated order flow, provided that market impact and liquidity responses compensate these correlations precisely~\citep{LilloFarmer2004,LilloMikeFarmer2005,BouchaudKockelkorenPotters2006}.

Market impact is the causal response of prices to trades -- the difference between realized and counterfactual prices~\citep{webster2023handbook} that would prevail in the absence of a given trading strategy. Hence diffusivity, as a structural property of both realized and counterfactual prices, imposes strong \emph{a priori} constraints on impact itself. 

This constraint operates at the level of individual market participants, and is therefore more restrictive than earlier balancing arguments based on aggregate order flow~\citep{BouchaudKockelkorenPotters2006, BouchaudFarmerLillo2008:how_markets_digest} or metaorder flow~\citep{maitrier2025subtle, sato2025brownian}. Under weak, empirically relevant assumptions, the resulting consistency condition essentially determines how market impact scales with time and traded volume. Two regimes emerge. When trading is uncorrelated with underlying price movements -- the information-neutral regime -- diffusivity forces impact increments to be effectively white, implying square-root scaling of its typical magnitude. When trading is instead aligned with underlying returns, impact grows linearly with volume. This reproduces both the square-root law of metaorder impact~\citep{AlmgrenThumHauptmannLi2005,GabaixGopikrishnanPlerouStanley2006, ChackoJurekStafford2008, moro2009markethidden,ZarinelliEtAl2015,Bacry2015MarketImpactLifeCycle,BucciBenzaquenLilloBouchaud2019,SatoKanazawa2025} and the recently observed linear-to-sublinear crossover~\citep{BucciBenzaquenLilloBouchaud2019}, thereby recovering the empirical phenomenology. 

The square-root law has often been regarded as anomalous~\citep{BouchaudKockelkorenPotters2006, MastromatteoTothBouchaud2014PRL}. Our approach suggests rather the opposite: square-root impact is not anomalous but a direct consequence of price diffusivity, independent of specific assumptions about aggregate demand and supply dynamics. More precisely, when trading is uncorrelated with underlying price movements, impact itself must be diffusive -- a structural requirement that models based on exogenous, universal kernels generally violate, raising the possibility that many propagator and latent liquidity models mischaracterize market impact.

\section{Structural Price Diffusivity as a Constraint}

Consider an investor trading in a market. Let $(x_t)$ denote demeaned returns measured over time bins of size $\Delta t$. We define the impact of the investor's trades, $j_t$, as the difference between the realized return and the counterfactual return $y_t$ that would have prevailed in the absence of trading~\citep{AlmgrenChriss2001, webster2023handbook}:
\be
\label{eq:decomposition}
x_t = y_t + j_t \; .
\ee

Persistent return autocorrelations are expected to be arbitraged away. This condition is enforced adaptively: arbitrageurs continuously calibrate to the statistical composition of order flow, restoring diffusivity when participants enter or exit the market. This economic mechanism is expected to operate in mature markets under any reasonable scenario. The counterfactual price $(y_t)$, corresponding to the scenario where the investor does not trade, is therefore also expected to be diffusive -- a fortiori so when the investor is large. Structural price diffusivity thus imposes constraints on both $(x_t)$ and $(y_t)$.  Sec.~\ref{sec:bound} derives a quantitative detectability criterion that makes this intuition precise.

Accordingly, we impose
\begin{align*}
\label{eq:realized}
R_{xx}[t] &= \Delta t \, \sigma_X^2 \, \delta[t] \\
R_{yy}[t] &= \Delta t \, \sigma_Y^2 \, \delta[t] \;,
\end{align*}
where $R_{ab}[t] = \mathbb{E}[a_{t+k} b_k]$ for any centered processes $a,b$, $\delta$ is the Kronecker delta, and $\sigma_X, \sigma_Y$ denote the volatilities of $(x_t)$ and $(y_t)$. While $\sigma_X$ is observable, $\sigma_Y$ is not. 

This constraint holds in an ex-post setting where trading activity is stationary, so that arbitrage can enforce diffusivity. It does not apply to transient interventions or point-in-time optimization (see discussion).

Under these assumptions, \eq{eq:decomposition} implies
\begin{equation}
\label{eq:fundamental}
R_{yj}[t] + R_{jy}[t] + R_{jj}[t]
= \Delta t \left(\sigma_X^2 - \sigma_Y^2\right) \delta[t] \;. 
\end{equation}
This constraint operates at the level of individual trading strategies. 

Note that the relevant object is not an individual execution episode -- such as a metaorder -- but the trading strategy as a stationary stochastic process. Adaptation need not occur within a single metaorder; it will, however, occur over the lifetime of any persistent, stationary strategy, as arbitrageurs respond to its unconditional flow statistics. Stationarity is an empirically relevant idealization: typical investors make only rare or incremental changes to their trading envelope -- deployed capital, risk aversion, informational alpha -- so that the statistics of their order flow remain stable over macroscopic horizons.

\section{Information-neutral benchmark}
\label{sec:neutral}

A participant's informational alpha enters \eq{eq:fundamental} through the coupling $R_{jy}$. In the information-neutral regime,
\bee
R_{jy}[t]= 0 \,, \qquad \text{for all $t$}\,.
\eee

A technical subtlety is that both order flow and counterfactual returns may depend on common market-state variables, such as liquidity or volatility. Formally, order flow carries no information about $(y_t)$ if $q \perp y|\mathfrak h$, where $\mathfrak h$ denotes all non-directional market states. Since $j$ is determined by $q$ and $\mathfrak h$, this implies $\ex{j_t|y, \mathfrak h}=\ex{j_t|\mathfrak h}$, which vanishes when $j$ is centered conditional on $\mathfrak h$; therefore $\ex{j_t|y}=0$ and $R_{jy}=0$.

This regime is relevant in practice, as a substantial fraction of institutional trading -- such as risk rebalancing, hedging, or long-horizon allocation -- is not or only weakly driven by short-term return forecasts (see e. g. the analysis of proprietary trades and signals in \citep{Bucci2019SlowDecay}).
\eq{eq:fundamental} then reduces to 
\begin{equation}
\label{eq:fundamental2}
R_{jj}[t] = (\sigma_X^2 - \sigma_Y^2) \,\Delta t \, \delta[t]\,.
\end{equation} 

Thus, structural diffusivity implies that impact increments are white for every information-neutral participant: the impact mechanism acts as a whitening transformation of individual order flow.  

Earlier market impact models enforced price diffusivity through a balance between aggregate order-flow correlations and impact decay~\citep{BouchaudFarmerLillo2008:how_markets_digest}. Diffusivity was then guaranteed only for the realized price under the flow statistics observed in the market. By contrast, counterfactual diffusivity requires invariance under changes in order flow composition: this constrains impact at the level of individual participants rather than only at the aggregate level. 
The second-order statistics of impact are therefore independent of the temporal fine-structure of the underlying trading strategy and fully characterized by their instantaneous variance. The resulting Brownian scaling of information-neutral impact provides the key mechanism from which the square-root impact law emerges (see below). 

Note that \emph{any} mechanical impact model compatible with structural diffusivity must satisfy the requirement \eq{eq:fundamental2}. But clearly, this requirement cannot, in general, be satisfied by models with universal, strategy-agnostic impact kernels.

\paragraph{Impact whitening and market adaptation. }

Market adaptability is key to this mechanism. Consider a market that is initially diffusive, with returns generated by both fundamentals and the aggregate flow of all participants except the participant under consideration. Let the participant introduce an additional order-flow component $(q_t)$. Let 
\bee
j^\text{raw} = G(q)
\eee
denote the immediate mechanical price response to $(q_t)$, determined by the prevailing liquidity environment. In general, $j^\text{raw}$ need not satisfy \eq{eq:fundamental}; indeed, if $(q_t)$ is autocorrelated, it will generally induce predictable returns,
\bee
x_t^\text{raw} = y_t + G(q)_t\;.
\eee

Such predictability creates statistical arbitrage opportunities. Liquidity providers, market makers, and arbitrageurs therefore adapt their behavior (e. g. modifying quotes) in response to the return patterns generated by $(G(q)_t)$. We denote the resulting equilibrium adjustments by $(K(G(q))_t)$, so that realized returns become:  
\bee
x_t = y_t + G(q)_t + K(G(q))_t\;.
\eee

Both $(G(q)_t)$ and $(K(G(q))_t)$ are causal consequences of the participant's trading activity: the former is the direct mechanical response, while the latter arises indirectly through the adaptive reaction of the rest of the market. Their sum therefore constitutes the total impact, 
\bee
j_t = G(q)_t + K(G(q))_t \;.
\eee

Since the background process $(y_t)$ is already diffusive, any effect of adaptation to a new strategy is marginal to that strategy. This need not happen because market participants explicitly detect individual participants, but because competitive arbitrage continuously removes the return predictability that such flows would otherwise generate.

\paragraph{The all-pass degree of freedom versus order flow surprise. }

Note that \eq{eq:fundamental2} does not imply that impact simply equals order flow surprise as, e. g., in~\citep{FarmerGerigLilloMike2006, FarmerGerigLilloWaelbroeck2013, Jaisson2015MarketImpact}. 

Assume $(q_t)$ is generated from white noise $\hat\eta$ through a causal, linear time-invariant (LTI) filter $L$,
\bee
q_t = (L\cdot \hat\eta)_t \;, \qquad \hat\eta \text{ standardized white Gaussian}\;,
\eee
where $\cdot$ denotes convolution. Then, \eq{eq:fundamental2} only implies that 
\be
\label{eq:jU}
j_t \propto (U\cdot\hat\eta)_t\,, \qquad \sum_{h\ge0}U[h]U[t+h] = \delta[t] \, \text{ for all $t\ge0$}\,.
\ee

The causal all-pass (orthogonal) filter $U$ is a degree of freedom not determined by structural price diffusivity. A simple example is
\be
\label{eq:allpass}
U[t] = u_0 \delta[t] - (1-u_0^2) u_0^{t-1} \mathds{1}_{t \ge 1} \,, \qquad 0<u_0<1\,,
\ee
i. e. an exponential propagator with a certain additional instantaneous component. Therefore, impact parametrized by \eq{eq:jU} whitens the flow and at the same time couples flow surprises in a nontrivial way, thereby allowing for transient effects such as impact decay -- see Sec.~\ref{sec:decay}.

\paragraph{Positive impact costs.}
\label{sec:costs}

Define impact costs as minus the impact markout on the trade,
\be
\mathcal{C} = -\ex{q_t \sum_{h\ge 1}j_{t+h}}\,.
\ee

Note that a pure order-flow surprise model $j_t \sim \hat\eta_t$, with $(\hat\eta_t)$ white, has manifestly zero impact costs. Fortunately, structural diffusivity leaves an all-pass degree of freedom which permits strictly positive costs, consistent with the no-arbitrage principle. As an example, assume that $(q_t)$ is generated through a biexponential filter $L$,
\be
\label{eq:biexponential}
q_t = (L\cdot \hat\eta)_t \;, \qquad L[t] = (1-z_1) z_1^t - (1-z_2) z_2^t\;, \qquad 0<z_1<z_2<1\;,
\ee
chosen so that $L[0]>0$ and $\sum_{t\ge 0} L[t] = 0$, which is necessary for the accumulated inventory $Q_t = \sum_{\ell \ge 0} q_{t-\ell}$ to be a stationary process~\footnote{This means that the strategy decomposes into many ``roundtrips''.}. With \eq{eq:allpass}, we find after some algebra
\bee
\mathcal{C} \propto -\sum_{t\ge0, h\ge 1}L[t] U[t+h] = (1-u_0^2)\frac{z_2-z_1}{(1-z_1 u_0)(1-z_2 u_0)} > 0
\eee
for $0\le z_1<z_2<1$ and $0\le u_0<1$. Therefore, any nontrivial phase of the form \eq{eq:allpass} generates positive costs under a biexponential trading process, consistent with the absence of price manipulation in the sense of \citet{HubermanStanzl2004}. This result is not limited to first-order all-passes: for an all-pass of any given order, there exists a parameter region that generates positive costs under the biexponential $L$. Furthermore, the general proposition that there exists an all-pass that guarantees positive costs for \emph{any} $L$ (not necessarily biexponential) also holds. A proof will be published elsewhere~\citep{bonart2026foundation}.

\section{Participation-rate scaling of impact volatility}

We now examine the scaling of the zero-lag volatility factor $\sigma_X^2 - \sigma_Y^2$. 

Modern execution brokers determine single order placement and size as a function of locally available liquidity. Taking liquidity selectively implies that single-trade volume is only weakly correlated with target participation ratio, and single-trade impact only weakly correlated with size~\citep{BouchaudBonartDonierGould2018}. Since price is close to a random walk, an empirical relationship~\citep{BouchaudFarmerLillo2008:how_markets_digest, BouchaudBonartDonierGould2018} between the number of market transactions and volatility is therefore
\be
\label{eq:sigmadiff}
\sigma_X^2\, \Delta t \simeq \lambda_X^2 N\,,
\ee
where $\lambda_X$ is the volatility per trade (of the order of the bid-ask spread) and $N$ the number of market transactions within the interval $\Delta t$. 
This relationship is very robust both on small-tick~\citep{WyartBouchaudKockelkorenPottersVettorazzo2008} and large-tick assets, provided a continuous proxy of the true equilibrium price is used~\citep{BouchaudBonartDonierGould2018, BonartLillo2018}. 

Consider a strategy with a participation rate $\mu$, i.e. $\Delta N = \mu N$ trades in the market are due to the strategy. Counterfactual price diffusivity then implies 
\bee
\sigma_Y^2\, \Delta t \simeq \lambda_Y^2 \, (N-\Delta N)\,.
\eee
Write
\bee
\lambda_X=\lambda(0)\,,\qquad \lambda_Y=\lambda(\mu) \,.
\eee
Taylor expand around small $\mu$:
\be
\label{eq:impactvol}
(\sigma_X^2 - \sigma_Y^2) \Delta t \simeq \left(\lambda^2(0) - \frac{\partial\lambda^2}{\partial \mu}(0) \right) \Delta N = \lambda_\text{imp}^2 N \mu = \mathcal{O}(N\mu) \,,
\ee
where $\lambda_\text{imp}$ measures the typical single-trade impact of the strategy. Thus, $\sigma_X^2 - \sigma_Y^2$ scales \emph{linearly} in the participation rate (at fixed $N$). 

This result can be further understood as follows: assume price dynamics are driven by many heterogeneous strategies, so total order flow can be decomposed into many orthogonal sub-flows with individual $\mu_r$, each contributing independently to total volatility with $\sigma_r^2$. Therefore,
\bee
\lambda_X^2 N = \sum_r \sigma_r^2 = \lambda_X^2 \sum_r (N \mu_r)\,.
\eee
Each orthogonal sub-flow only ``sees'' the aggregate rest, so $\sigma_r^2 = \sigma_r^2\left(\mu_r, 1-\mu_r\right)$ which is thus a function of $\mu_r$ only. Write $\sigma_r^2 = S(\mu_r)$. Since prices should be structurally diffusive irrespective of the decomposition $\{\mu_r\}$, $S$ must be linear. 

Finally, combining \eq{eq:sigmadiff} and \eq{eq:impactvol} yields
\be
\label{eq:impactvol2}
\sigma_X^2 - \sigma_Y^2 \simeq \Lambda^2\, \sigma_X^2\, \mu\,, \qquad \Lambda = \frac{\lambda_\text{imp}}{\lambda_X}\,,
\ee
where $\Lambda$, the ratio of strategy-level single-trade impact to market-wide single-trade impact, captures execution quality.

\section{Consequences of the structural constraint}

\eq{eq:fundamental}, \eq{eq:fundamental2} and \eq{eq:impactvol} impose nontrivial constraints on how impact scales with trading activity. Under mild and empirically relevant assumptions, these constraints imply a diffusive impact in an information-neutral setting, a square-root scaling in duration, a linear-to-square-root crossover of impact as a function of trading rate and informational alpha and, consequently, the empirical square-root law of metaorder impact.

\paragraph{Double square-root scaling of informationless impact. }

For calendar time $\tau=t\Delta t$, define the impact increment over a horizon of $h$ days by
\bee
I_\tau = \sum_{k=(\tau-h)/\Delta t+1}^{\tau/\Delta t} j_{k}\,, \qquad \tau = t\Delta t \,.
\eee
Write 
\bee
I_\tau = I_\tau(h, \mu)
\eee
to make the dependence on both horizon $h$ and participation rate $\mu$ explicit. In the hydrodynamic limit, $\Delta t\to 0$, assuming the central limit theorem applies, \eq{eq:fundamental2} and \eq{eq:impactvol} imply
\be
\label{eq:scalingY}
I_\tau(h, \mu) \overset{d}{=} \sqrt{h \, (\mu/\mu_0)} \, I_\tau(1, \mu_0) \,, \qquad 
\std{I_\tau(1, \mu_0)} = \lambda_\text{imp}\sqrt{N_D \mu_0}\;,
\ee
where $\mu_0$ is an arbitrary reference participation rate and $N_D$ is the daily number of market transactions.

Assuming that metaorders are typical signed excursions of stationary trading activity, conditioning on their direction and trading rate $\mu$ over the calendar-time interval $[\tau-h,\tau]$ fixes the sign and prefactor but leaves the scaling exponent unchanged (see below):
\be
\label{eq:scalingYm}
I_\tau(h, \mu) | \text{metaorder} \propto \sqrt{h \, (\mu/\mu_0)} \, I_\tau(1,\mu_0) | \text{metaorder}\;,
\ee
where the proportionality constant does not depend on $\mu$ and $h$. 
We discuss metaorder conditioning more rigorously below. 

$I_\tau(1,\mu_0)| \text{metaorder}$ is positive for a buy and negative for a sell metaorder. Matching to single-trade impact, \eq{eq:scalingY} implies
\bee
I_\tau(h, \mu) | \text{metaorder} \propto \text{sign}(Q)\, \lambda_\text{imp} \, \sqrt{\frac{|Q|}{\nu}}\;,
\eee
where $\nu$ is the typical trade size, $Q$ the total volume of the metaorder, and thus $|Q|/\nu$ is the number of executions from a metaorder; the square-root law in duration and trading rate is essentially a square-root law in number of child orders, consistent with \citep{Lodhia2026ImpactModels, Vasaikar2026SquareRoot}. 

Daily volatility scales as $\sigma_X \sim \lambda_X\sqrt{N_D}$, daily volume as $V \sim \nu N_D$, and $\lambda_\text{imp} \approx \lambda_X$ if strategy executions are market typical. Then,
\bee
I_\tau(h, \mu) | \text{metaorder} \propto \text{sign}(Q) \, \sigma_X \, \sqrt{\frac{|Q|}{V}}\;.
\eee
This is the canonical square-root law~\citep{BouchaudBonartDonierGould2018}.

This scaling is valid until (unmodeled) restoring forces set in. We cannot pin down the precise horizon when this happens; trend and mean-reversion effects appear insufficient, over horizons shorter than several weeks, to invalidate the conclusions drawn from structural price diffusivity.
To quantify this intuition, consider a trader executing $1$--$2\%$ of daily volume. This corresponds to a weekly impact level of the order of $25\%$ of volatility, and a quarterly impact level around $100\%$, since $\sqrt{63} \times \sqrt{0.02} \approx 1$. The result is plausible: the mechanical mispricing caused by this trading strategy can reach $50$--$200$ basis points for a typical stock ($2\%$ daily volatility) until major deviations from theory are expected. This is also where one would expect the impact phase $U$ to break down, i.e. with departures from price diffusivity, impact decay need not be constrained by an all-pass dynamics.

Sec.~\ref{sec:bound} derives a high-frequency bound on the regime of validity.

\paragraph{Metaorder impact diffusivity. }
\label{sec:decay}

The transition from \eq{eq:scalingY} to \eq{eq:scalingYm} requires that metaorders generate the typical fluctuations of impact.
 
Consider a strategy consisting of metaorders with fixed duration and trading rate $(h,\mu)$, and choose the bin length $\Delta t = h$. Assume metaorder signs $(\epsilon_t)$ are generated by a ternary discretization of an autocorrelated latent Gaussian process $(\eta_t)$ via a causal LTI filter,
\bee
\eta_t = (\tilde L \cdot \hat\eta)_t \;, \qquad \hat\eta_t \sim \text{ standardized white Gaussian} \,,
\eee
so that
\bee
\epsilon_t = 
\begin{cases}
+1 \,, & \eta_t > a \\
-1 \,,& \eta_t < -a \\
0 \,,& \text{else}
\end{cases}\;,
\eee
where $a>0$ is some threshold. 
Impact is diffusive when
\bee
j_t = A(h, \mu)\,(\tilde U\cdot \hat\eta)_t\,,
\eee
where $\tilde U$ is again a causal all-pass, and
\bee
A(h, \mu) \sim \sqrt{\mu h}
\eee  
is the impact amplitude fixed by impact diffusivity. 

Applying the Gaussian regression formula, we find
\bee
\ex{j_t|\eta_t} = \eta_t\,\frac{\mathrm{Cov}(j_t, \eta_t)}{\mathrm{Var}(\eta_t)} = \eta_t\,A(h, \mu)\,\frac{\sum_{k\ge 0} \tilde U[k]\,\tilde L[k]}{\sum_{k\ge 0} \tilde L[k]^2}\,,
\eee
and, using $\ex{\eta_t|\epsilon_t} \propto \epsilon_t$, we recover the square-root scaling of average metaorder impact,
\bee
\ex{j_t|\epsilon_t} \propto \epsilon_t \, \sqrt{\mu h}\,.
\eee
Note that the prefactor depends on the undetermined all-pass filter $\tilde U$ and on the parametrization $(a, \tilde L)$ of the metaorder sign process, but not on $\mu$ or $h$.

A nontrivial $\tilde U \ne I$ encodes a departure from a pure surprise mechanism, in which impact responds only to the contemporaneous order flow innovation. It therefore allows for nontrivial impact decay. For simplicity, consider a flow of uncorrelated metaorders. The lag-$1$ post-order response then satisfies
\bee
\ex{j_t | \epsilon_t = 0, \epsilon_{t-1}} \sim \sqrt{\mu h}\, \tilde U[1]\, \epsilon_{t-1} \,.
\eee
Since slippage is positive, $\tilde U[0]>0$, but $\tilde U[1]$ may be negative, as is the case for the simple exponential all-pass filter in \eq{eq:allpass}. 

While structural price diffusivity alone does not determine $\tilde U$, it remains fully compatible with the empirically observed relaxation of informationless impact~\citep{gomes2015market, brokmann2015slow, Bucci2019SlowDecay, FarmerGerigLilloWaelbroeck2013}, despite its diffusivity.


\paragraph{Informational coupling: Linear-to-square-root crossover.}

$R_{yj}$ encodes the coupling between impact and counterfactual returns: positive lags generally correspond to forward predictability and negative lags to endogeneity -- trades induced by price moves, e. g. passive executions. 

Mechanical impact dominates informational alpha in the regime
\bee
R_{jj}[0] \gg 2|R_{jy}[0]|\;.
\eee
\eq{eq:impactvol2} then implies
\be
R_{jj}[0] = \Lambda^2\, \sigma_X^2 \, \mu\, \Delta t  \qquad \text{strong impact, weak information.}
\ee
Thus, the typical impact magnitude scales as $\mathcal{O}(\sigma_X \sqrt{\mu})$. 

Conversely, in the weak impact regime
\bee
R_{jj}[0] \ll 2|R_{jy}[0]|
\eee
we have
\be
R_{yj}[0] = \frac{1}{2} \Lambda^2\, \sigma_X^2 \, \mu\, \Delta t  \qquad \text{weak impact with information.}
\ee
Since $y$ is independent of $\mu$ with a volatility roughly equal to $\sigma_X$, and assuming a fixed correlation structure with the counterfactual return, the typical impact magnitude must now scale as $\mathcal{O}(\mu \sigma_X)$. 

Altogether, therefore,
\be
\label{eq:impactscaling}
j \sim \sigma_X \, \sqrt{\Delta t} \, \varphi(\mu) \,, \qquad \varphi(\mu) = \begin{cases}
 \mu & \text{weak impact with information} \\
\sqrt{\mu} & \text{strong impact, weak information}
\end{cases}
\ee

The crossover occurs when $j \sim \hat y$ in magnitude, where $\hat y$ is the contemporaneous informational alpha, i.e. the $y$-component that is correlated with the strategy. Thus, impact is more linear for strategies that trade on relatively stronger short-term alpha. 
This linear departure from the square-root law is indeed consistent with empirical findings~\citep{ZarinelliEtAl2015,BucciBenzaquenLilloBouchaud2019}.

\section{A bound on short-time diffusivity}
\label{sec:bound}

A nontrivial
\bee
\mathcal{R}[t] = R_{yj}[t] + R_{jy}[t] + R_{jj}[t]
\eee
must be detectable for diffusivity to be restored by arbitrageurs. Applying the Kolmogorov-Szeg\H{o} formula~\citep{Palma2016:TimeSeriesAnalysis} for small perturbations, the optimal predictor under the given covariance $R_{xx}[t] = \sigma_Y^2\Delta t\delta[t] + \mathcal{R}[t]$ satisfies
\bee
\std{\hat x} \simeq \frac{1}{\sigma_Y \sqrt{\Delta t}} \left(\sum_{t\ge 1} \mathcal{R}[t]^2\right)^{1/2}\,, \qquad \frac{\std{\hat x}}{\sigma_Y\sqrt{\Delta t}} \ll 1\,.
\eee
Choose the bin length equal to the horizon, measured in days, over which impact remains aligned:
\be
\Delta t = \frac{n_\text{coh}}{\mu N_D}\,,
\ee
where $n_\text{coh}$ is the number of strategy-executions over which impact remains order-one correlated and $N_D$ is the number of market transactions per day. Then, a rough triangular approximation of the double sum gives
\bee
\mathcal{R}[h] \sim \left(\lambda^2 + 2\lambda\alpha\right)n_\text{coh}^2 \;, \qquad h \le 1
\eee
and $\mathcal{R}[h] \approx 0$ for larger horizons. Here, $\lambda$ denotes the impact scale of a typical trade and $\alpha$ the coupling to underlying price moves. Assuming $\lambda_X \approx \lambda_\text{imp} \approx \lambda$, and using again the empirical scaling
\bee
\sigma_Y \approx \sigma_X \sim \lambda \sqrt{N_D},
\eee
this simplifies to
\bee
\std{\hat x} \sim (\lambda + 2\alpha) \sqrt{\mu \, n_\text{coh}^3} \,.
\eee

Empirically, $\lambda$ is an order of magnitude smaller than the bid-ask spread~\citep{BouchaudFarmerLillo2008:how_markets_digest}, denoted $S$, which is also the threshold beyond which arbitrage, liquidity response and impact decay start to blunt predictable patterns in the order flow. Mid-price diffusivity sets in when the volatility exceeds the spread, and continuous proxies of the equilibrium price that smooth high-frequency bid-ask bounces are diffusive much earlier, as demonstrated by accurate predictions of the Madhavan-Richardson-Roomans model~\citep{MadhavanRichardsonRoomans1997} on large-tick assets~\citep{BouchaudBonartDonierGould2018, BonartLillo2018}.

Thus, impact autocorrelations are suppressed beyond
\be
n_\text{coh}^* \sim C\mu^{-1/3} \qquad C = \left(\frac{S}{\lambda+2\alpha}\right)^{2/3}
\ee
and structural price diffusivity, \eq{eq:fundamental}, operates for bin sizes $\gtrsim n_\text{coh}^*$, measured in execution clock. 
With the choice $C \sim 1$, we find $n_\text{coh}^* \approx 5$ executions for a moderate participation ratio $\mu = 1\%$, and $n_\text{coh}^* \approx 2$ for $\mu = 10\%$.
More specifically, we therefore expect impact of a $1\%$ strategy executed on a stock with (say) $20,000$ daily transactions to be whitened after ten minutes. In other words, structural diffusivity is an accurate description of the impact dynamics induced by typical multiday metaorders.

\section{Discussion}

\paragraph{On optimal execution and price manipulation.}

Structural price diffusivity suggests that much of the optimal execution literature neglects a central ingredient: the adaptive nature of impact. Our framework implies that the historical impact of stationary strategies should not be extrapolated to point-in-time optimization scenarios outside the historical statistical envelope of order flow. Within this envelope, the observed impact law may be useful for optimization over short horizons. But when a genuinely new trading regime is explored -- for example, through a fundamental change in trading style or deployed capital -- the previously inferred impact law may become useless.

The theory predicts that the impact law itself changes under persistent changes in trading behavior, potentially offsetting gains predicted by optimizing a strategy against an ex-ante calibrated impact kernel. One practical implication is that backtests or optimizers exploring scenarios outside this envelope should be treated with caution.

A theoretical implication concerns the standard notion of price manipulation~\citep{HubermanStanzl2004,Gatheral2010,AlfonsiSchied2010,GatheralSchiedSlynko2012}. Under structural price diffusivity, market impact is adaptive. Hence, the following two propositions may simultaneously hold: a fixed impact law admits a price-manipulating strategy, and that strategy lies outside the manifold of execution scenarios over which the impact law remains invariant. In this case, the apparent manipulation opportunity is not exploitable: the impact law that generates it shifts like a moving target.

That said, impact costs \emph{after} adaptation to a participant's stationary order flow should be non-negative. We show that these costs are indeed strictly positive under certain conditions (see Sec.~\ref{sec:costs}). A corresponding result for general LTI kernels is possible~\citep{bonart2026foundation}, but lies beyond the scope of this work.

\paragraph{The theory further predicts:}

\begin{enumerate}

\item[(1)] A square-root law primarily in the number of child orders. Consistent with our framework, \citet{Lodhia2026ImpactModels} finds a near-zero dependence on volume once trade count is controlled for. If further corroborated, this empirical finding would also provide a sharp demarcation between our framework and the latent-liquidity model~\citep{DonierEtAl2014}. 

\item[(2)] Any stationary sequence of trades should generate square-root impact in the information-neutral regime, irrespective of its origin. Consistently, recent work finds square-root scaling even for ``synthetic'' metaorders constructed from market order flow via a stationary selection procedure~\citep{MaitrierLoeperBouchaud2025, Vasaikar2026SquareRoot}.

\item[(3)] Stronger alignment between order flow and informational alpha reduces the concavity of impact. In particular, conditioning metaorder impact on the concurrent residual market imbalance should produce a linear departure from square-root scaling in trading rate.

\item[(4)] A rationale for, and generalization of, Market Microstructure Invariance. Decompose order flow into independent ``bets'', each contributing additively a volume $Q_b$ and variance $\sigma_b^2$, so that
\be
\label{eq:mmi}
\sigma V = \sigma_b Q_b N^{3/2}\,,
\ee
where $\sigma$ denotes volatility, $V$ the traded volume measured over a given interval, and $N = V/Q_b$ the number of independent bets. Market Microstructure Invariance is the hypothesis that the exchanged ``bet risk'' $\sigma_b Q_b$ is invariant across assets and time horizons~\citep{KyleObizhaeva2016}. Early empirical studies reported good agreement with this hypothesis~\citep{KyleObizhaeva2016, AndersenBondarenkoKyleObizhaeva2018}, whereas later work~\citep{BenzaquenDonierBouchaud2016, BucciLilloBouchaudBenzaquen2020} documented substantial deviations, consistent with the observation that $\sigma_b Q_b$ scales with the impact cost of a bet.

\eq{eq:mmi} is tautological under the abstract notion of independent bets, but becomes nontrivial when applied to metaorders or individual transactions, which are manifestly correlated.

Structural price diffusivity provides a natural rationale. Rewriting \eq{eq:mmi} yields
\be
\label{eq:mmi2}
\sigma_b = \sigma\frac{\sqrt{Q_b}}{\sqrt{V}}\;.
\ee
Under structural diffusivity, impact increments generated by any stationary strategy must be asymptotically uncorrelated, so their induced variances add up, even if the bets themselves remain correlated. Consequently, \eq{eq:mmi2} should hold across a broad range of horizons for any trading strategy sufficiently orthogonal to the rest of the market, and \emph{a fortiori} for the market as a whole.

\end{enumerate}

\section{Conclusion}

Let us retrace the main argument: if both the observed and counterfactual prices are diffusive, their difference -- market impact -- satisfies a structural constraint which, unlike earlier balancing arguments applied to aggregate flow, operates at the level of each individual participant. 

This constraint nearly fixes the admissible scaling structure of market impact. In the absence of informational coupling, impact increments must be effectively uncorrelated, independently of aggregate supply and demand dynamics. Information-neutral impact is therefore diffusive. Conversely, a square-root-to-linear crossover is predicted to occur when information content starts to dominate mechanical impact. 
Our framework thus recovers the empirical phenomenology. It further indicates that square-root impact is not anomalous and need not reflect market instability due to critically low liquidity, but instead arises as a generic consequence of adaptive price formation.

Consider the entry of a new trading strategy whose activity induces predictable price patterns. As long as this activity is stationary and persistent, arbitrage restores diffusivity. But the arbitrage trades themselves generate mechanical impact, leading to a cascade across time scales. The process stabilizes when the new equilibrium price recovers diffusivity, reflecting market impact that is specific to the new strategy, yet does not require explicit detection of the participant: only the aggregate statistical constraint must be satisfied. In fact, any stationary sequence of trades should generate square-root impact in the information-neutral regime, regardless of its origin. 

Under this interpretation, impact is not merely the response to aggregate flow. Rather, it is the equilibrium response of an adaptive market to the introduction of the flow component of a particular strategy into an already diffusive environment. Since diffusivity constrains the marginal contribution of each strategy, equilibrium adaptation is generally path-dependent: introducing two strategies sequentially, allowing the market to re-equilibrate after each addition, need not lead to the same impact decomposition as introducing their aggregate flow at once. Consequently, the impact of an aggregate strategy need not equal the sum of the impacts of its constituents, even when the latter are traded independently. Different histories by which the same aggregate flow is composed can therefore correspond to distinct, yet equally diffusive, market equilibria.

Impact diffusivity also implies that the impact magnitude at the single-trade level matches the volatility of impact fluctuations over larger horizons. Does this imply that long-horizon impact is set by microstructural scales? Probably not: the microstructural MRR model~\citep{MadhavanRichardsonRoomans1997} yields accurate predictions about the equilibrium price even on sub-tick scales~\citep{BonartLillo2018}. In other words, even substantial price discretization -- despite its mechanical constraints -- does not break diffusivity. Furthermore, profound changes in market structure -- e. g. the rise of high-frequency trading, tighter spreads, much lower latency, and increased volume -- have had only limited effect on long-run volatility and market stability. This suggests a top-down mechanism whereby long-term volatility sets microstructural scales such as single-trade impact. The key implication remains: impact must be diffusive in the informationless regime. 

Crucially, all these results hold for stationary trading activity and should be understood as an unconditional, ex-post property. They do not apply to abrupt interventions or point-in-time optimal execution problems, where deviations from diffusivity may persist over finite horizons. This bears some similarity with the premise in \citet{SaddierMarsili2024} who interpret the square-root law of single metaorders as an Bayesian posterior. 

Finally, the framework provides a diagnostic for mechanical impact models. In short, any model that fails to whiten order flow violates structural diffusivity. When used to infer de-impacted prices for simulation purposes, such models produce autocorrelated return trajectories, indicating a systematic inconsistency. 
For example, the latent liquidity model -- to which the author has contributed in the past~\citep{DonierEtAl2014} -- does not generally satisfy this constraint. 
This restrictiveness was not anticipated. Yet many models imply that scaling a strategy or removing it altogether would lead to inefficient, non-diffusive price dynamics, a conclusion difficult to reconcile with expectations and the empirical observation that prices quickly revert to diffusivity following major market events, structural changes, or regime shifts.

\section*{Disclosure of Interests}

There are no interests to disclose.

\section*{Declaration of Funding}

No funding was received.

\section*{Generative AI statement}

OpenAI language models were used during the preparation of this manuscript to assist with proofreading.

\section*{Acknowledgements}

The author thanks Nicholas Westray, Johannes Muhle-Karbe, Xavier Gabaix, Natascha Hey, Guillaume Maitrier, Jean-Philippe Bouchaud, Iacopo Mastromatteo and Bence Toth for the interesting and helpful comments. 
The author acknowledges fruitful discussions with Galin Georgiev. The author also wishes to thank the two anonymous referees for their critical reading of the manuscript.

\bibliographystyle{plainnat}
\bibliography{square-root-impact}

\end{document}